\documentclass{elsart}
\usepackage{amssymb, latexsym, euscript}

 \newtheorem{ttt}{\bfseries{Theorem}}[section]
 
 \newtheorem{ppp}{\bfseries{Proposition}}[section]
 \newtheorem{lelele}{\bfseries{Lemma}}[section]

\journal{}
\begin{document}
\begin{frontmatter}
\title{$p$-Adic Schr\"{o}dinger-Type Operator with Point Interactions}
\author[f,g,r]{S. Albeverio}
\ead{albeverio@uni-bonn.de}
\author[s]{S. Kuzhel\corauthref{cor}}
\corauth[cor]{Corresponding author.}
\ead{kuzhel@imath.kiev.ua}
\author[s]{S. Torba}
\ead{sergiy.torba@gmail.com}
\address[f]{Institut f\"{u}r Angewandte Mathematik, Universit\"{a}t Bonn, Wegelerstr. 6, D-53115 Bonn (Germany)}
\address[g]{SFB 611, Bonn, BiBoS, Bielefeld-Bonn}
\address[r]{CERFIM, Locarno and USI (Switzerland)}
\address[s]{Institute of Mathematics of the National Academy of
Sciences of Ukraine, Tereshchenkovskaya 3, 01601 Kiev (Ukraine)}

\begin{abstract}
A $p$-adic Schr\"{o}dinger-type operator $D^{\alpha}+V_Y$ is
studied. $D^{\alpha}$ ($\alpha>0$) is the operator of fractional
differentiation and $V_Y=\sum_{i,j=1}^nb_{ij}<\delta_{x_j},
\cdot>\delta_{x_i}$  $(b_{ij}\in\mathbb{C})$ is a singular potential
containing the Dirac delta functions $\delta_{x}$ concentrated on a
set of points $Y=\{x_1,\ldots,x_n\}$ of the field of $p$-adic
numbers $\mathbb{Q}_p$. It is shown that such a problem is
well-posed for $\alpha>1/2$ and the singular perturbation $V_Y$ is
form-bounded for $\alpha>1$. In the latter case, the spectral
analysis of $\eta$-self-adjoint operator realizations of
$D^{\alpha}+V_Y$ in $L_2(\mathbb{Q}_p)$ is carried out.
\end{abstract}
\end{frontmatter}
 \section{Introduction}
The non-Archimedean analysis based on $p$-adic numbers  has a long
history and a quite exhaustive presentation of its applications in
stochastics, psychology, the theory of dynamical systems, and other
areas can be found in \cite{KH}, \cite{KH1}, \cite{KO}, \cite{VVZ}.
A strong impetus to the development of $p$-adic analysis was given
by the hypothesis about a possible $p$-adic structure of physical
space-time at sub-Planck distances ($\leq{10^{-33}}$ cm) \cite{VVZ}.
This idea gave rise to many publications (see the surveys in
\cite{KO}, \cite{VVZ}). Whatever form the $p$-adic models may take
in the future, it has become clear that finding $p$-adic
counterparts for all basic structures of the standard mathematical
physics is an interesting task.

In the present paper we are going to continue the investigation of
$p$-adic Schr\"{o}dinger-type operators with point interactions
started by A. Kochubei \cite{KO}.

In `usual' mathematical physics Schr\"{o}dinger operators with point
interactions are well-studied  and they are used in quantum
mechanics to obtain Hamiltonians describing realistic physical
systems but having the important property of being exactly solvable,
i.e., that all eigenfunctions, spectrum, and scattering matrix can
be calculated  \cite{AL}, \cite{AL1}.

Since there exists a $p$-adic analysis based on the mappings from
$\mathbb{Q}_p$ into $\mathbb{Q}_p$ and an analysis connected with
the mapping $\mathbb{Q}_p$ into the field of complex numbers
$\mathbb{C}$, there exist two types of $p$-adic physical models. The
present paper deals with the mapping $\mathbb{Q}_p\to{\mathbb C}$,
i.e., complex-valued functions defined on $\mathbb{Q}_p$ will be
considered. In this case the operation of differentiation  {\it is
not defined} and the operator of fractional differentiation
$D^{\alpha}$ of order $\alpha$ ($\alpha>0$) plays a corresponding
role \cite{KO}, \cite{VVZ}. In particular, $p$-adic
Schr\"{o}dinger-type operators with potentials $V(x) :
\mathbb{Q}_p\to\mathbb{C}$ are defined as $D^{\alpha}+V(x)$.

The definition of $D^{\alpha}$ is given in the framework of the
$p$-adic distribution theory with the help of Schwartz-type
distributions $\mathcal{D}'(\mathbb{Q}_p)$. One of remarkable
features of this theory is that any distribution
$f\in\mathcal{D}'(\mathbb{Q}_p)$ with point support $\mathrm{supp}
f=\{x\}$ coincides with the Dirac delta function at the point $x$
multiplied by a constant $c\in\mathbb{C}$, i.e., $f=c\delta_{x}$.

For this reason, it is natural to consider the expression
$D^{\alpha}+V_Y$ where the singular potential
$V_Y=\sum_{i,j=1}^nb_{ij}<\delta_{x_j}, \cdot>\delta_{x_i}$
$(b_{ij}\in\mathbb{C})$ contains the Dirac delta functions
$\delta_{x}$ concentrated on points $x_k$ of the set
$Y=\{x_1,\ldots,x_n\}\subset\mathbb{Q}_p$ as a $p$-adic analogue of
the Schr\"{o}dinger operator with point interactions.

Since $D^{\alpha}$ is a $p$-adic pseudo-differential operator the
expression $D^{\alpha}+V_Y$ gives an example of pseudo-differential
operators with point interactions. In the `usual' (Archimedean)
theory, expressions of such (and more general) type have been
studied in \cite{AK}.

Obviously the domain of definition $\mathcal{D}(D^{\alpha})$ of the
unperturbed operator $D^{\alpha}$ need not contains functions
continuous on $\mathbb{Q}_p$ and, in general, may happen that the
singular potential $V_Y$ is not well-defined on
$\mathcal{D}(D^{\alpha})$.

In Section 2, together with a presentation of some elements of
$p$-adic analysis needed for reading the paper, we discuss the
problem of characterizing $\mathcal{D}(D^{\alpha})$ and study in
detail  the solutions of the equation
$D^{\alpha}-\lambda{I}=\delta_x$.

Section 3 deals with the spectral analysis of operator realizations
of $D^{\alpha}+V_Y$ ($\alpha>1$) in $L_2(\mathbb{Q}_p)$.  We do not
restrict ourselves only to the self-adjoint case and also consider
$\eta$-self-adjoint operators. The investigation of such operators
is motivated by an intensive development of pseudo-Hermitian
($\mathcal{PT}$-symmetric) quantum mechanics in the last few years
\cite{BBJ}, \cite{GF},  \cite{MO}, \cite{TT}, \cite{ZN}.

Among self-adjoint extensions of the symmetric operator
$A_{\mathrm{sym}}$ associated with $D^{\alpha}+V_Y$ $(\alpha>1)$, we
pay a special attention to the Friedrichs extension $A_F$. Since
$A_F$ is the `hard' extension of $A_{\mathrm{sym}}$ (see \cite{AN}
for the terminology) and the singular potential $V_Y$ is form
bounded the hypothesis that the discrete spectrum of $A_F$ depends
on the geometrical structure of $Y$ looks likely. In this way we
discuss the connection between the minimal distance
$p^{\gamma_{\mathrm{min}}}$ between elements of $Y$ and an infinite
sequence of points of the discrete spectrum (type-$1$ part of
discrete spectrum).

We will use the following notations: $\mathcal{D}(A)$ and $\ker{A}$
denote the domain and the null-space of a linear operator $A$,
respectively. $A\upharpoonright_{X}$ means the restriction of $A$
onto a set $X$.

\setcounter{equation}{0}
\section{Fractional Differential Operator}
\subsection{Elements of $p$-adic analysis.}
Basically we shall use the same notations as in \cite{VVZ}. Let $p$
be a prime number.  The field $\mathbb{Q}_p$ of $p$-adic numbers is
the completion of the field of rational numbers $\mathbb{Q}$ with
respect to $p$-adic norm $|\cdot|_p$, which is defined as follows:
$|0|_p=0$; $|x|_p=p^{-\gamma}$ if a rational number $x\not=0$ has
the form $x=p^{\gamma}\frac{m}{n}$, where
$\gamma=\gamma(x)\in\mathbb{Z}$ and integers $m$ and $n$ are not
divisible by $p$. The $p$-adic norm $|\cdot|_p$ satisfies the strong
triangle inequality $|x+y|_p\leq\max(|x|_p, |y|_p)$. Moreover,
$|x+y|_p=\max(|x|_p, |y|_p)$ if $|x|_p\not=|y|_p$.

Any $p$-adic number $x\not=0$ can uniquely be presented as a series
\begin{equation}\label{e2}
x=p^{\gamma}\sum_{i=0}^{+\infty}x^ip^i, \qquad x^i=0,1,\ldots,p-1,
\quad x^0>0
\end{equation}
convergent in the $p$-adic norm (the canonical presentation of $x$).

The canonical presentation (\ref{e2}) enables one to determine the
fractional part $\{x\}_p$ of $x\in{\mathbb{Q}_p}$ by the rule:
$\{x\}_p=0$ if $x=0$ or $\gamma(x)\geq{0}$;
$\{x\}_p=p^{\gamma(x)}\sum_{i=0}^{-\gamma(x)-1}x^ip^i$ if
$\gamma(x)<{0}$.

Denote by $B_{\gamma}(a)=\{x\in\mathbb{Q}_p \ | \
|x-a|_p\leq{p^\gamma}\}$ the ball of radius $p^\gamma$ with the
center at a point $a\in\mathbb{Q}_p$ and set
$B_{\gamma}(0)=B_\gamma$.  The ring $\mathbb{Z}_p$ of $p$-adic
integers is the ball $B_0$ ($\mathbb{Z}_p=B_0$).

A complex-valued function $f$ defined on $\mathbb{Q}_p$ is called
{\it locally-constant} if for any $x\in\mathbb{Q}_p$ there exists an
integer $l(x)$ such that $f(x+x')=f(x)$, \
$\forall{x'}\in{B_{l(x)}}$.

Denote by $\mathcal{D}(\mathbb{Q}_p)$ the linear space of locally
constant functions on $\mathbb{Q}_p$ with compact supports. For any
test function $\phi\in\mathcal{D}(\mathbb{Q}_p)$ there exists
$l\in\mathbb{Z}$ such that $\phi(x+x')=\phi(x)$, \ $x'\in{B_{l}}$,\
$x\in\mathbb{Q}_p$. The largest of such numbers $l=l(\phi)$ is
called the {\it parameter of constancy} of $\phi$. The
characteristic function $\Omega(|x|_p)=1$ if $|x|_p\leq{1}$ and
$\Omega(|x|_p)=0$ if $|x|_p>{1}$ of the ball $B_0$ is an example of
test functions with parameter of constancy 1.

In order to furnish $\mathcal{D}(\mathbb{Q}_p)$ with a topology, let
us consider the subspace
$\mathcal{D}^l_\gamma\subset\mathcal{D}(\mathbb{Q}_p)$ consisting of
functions with supports in the ball $B_\gamma$ and the parameter of
constancy $\geq{l}$. The convergence $\phi_n\to{0}$ in
$\mathcal{D}(\mathbb{Q}_p)$ means that:
$\phi_k\in\mathcal{D}^l_\gamma$, where the indices $l$ and $\gamma$
do not depend on $k$ and $\phi_k$ tends uniformly to zero. This
convergence determines the Schwartz topology in
$\mathcal{D}(\mathbb{Q}_p)$.

Denote by $\mathcal{D}'(\mathbb{Q}_p)$ the set of all linear
functionals (Schwartz-type distributions) on
$\mathcal{D}(\mathbb{Q}_p)$. In contrast to distributions on
$\mathbb{R}$, any linear functional
$\mathcal{D}(\mathbb{Q}_p)\to\mathbb{C}$ is automatically
continuous. The action of a functional $f$ upon a test function
$\phi$ will be denoted as $<f, \phi>$.

It follows from the definition of $\mathcal{D}(\mathbb{Q}_p)$ that
any test function $\phi\in\mathcal{D}(\mathbb{Q}_p)$ is continuous
on $\mathbb{Q}_p$. This means that the Dirac delta function
$<\delta_x, \phi>=\phi(x)$ is well defined for any point
$x\in\mathbb{Q}_p$.

On $\mathbb{Q}_p$ there exists the Haar measure, i.e., a positive
measure $d_px$ invariant under shifts $d_p(x+a)=d_{p}x$ and
normalized by the equality $\int_{|x|_p\leq{1}}d_{p}x=1$.

Denote by $L_2(\mathbb{Q}_p)$ the set of measurable functions $f$ on
$\mathbb{Q}_p$ satisfying the condition
$\int_{\mathbb{Q}_p}|f(x)|^2d_{p}x<\infty$. The set
$L_2(\mathbb{Q}_p)$ is a Hilbert space with the scalar product
$(f,g)=\int_{\mathbb{Q}_p}f(x)\overline{g(x)}d_{p}x$.

The Fourier transform of $\phi\in\mathcal{D}(\mathbb{Q}_p)$ is
defined by the formula
$$
F[\phi](\xi)=\widetilde{\phi}(\xi)=\int_{\mathbb{Q}_p}\chi_p(\xi{x})\phi(x)d_{p}x,
\qquad \xi\in\mathbb{Q}_p,
$$
where $\chi_p(\xi{x})=e^{2\pi{i}\{\xi{x}\}_p}$ is an additive
character of the field $\mathbb{Q}_p$ for any $\xi\in\mathbb{Q}_p$.
The Fourier transform $F[\cdot]$ maps $\mathcal{D}(\mathbb{Q}_p)$
onto $\mathcal{D}(\mathbb{Q}_p)$. Its extension by continuity onto
$L_2(\mathbb{Q}_p)$ determines an unitary operator in
$L_2(\mathbb{Q}_p)$.

The Fourier transform $F[f]$ of a  distribution
$f\in\mathcal{D}'(\mathbb{Q}_p)$ is defined by the standard relation
$<F[f], \phi>=<f, F[\phi]>$, \
$\forall{\phi}\in\mathcal{D}(\mathbb{Q}_p)$.

\subsection{The operator $D^{\alpha}$.}
The operator of differentiation is not defined in
$L_2(\mathbb{Q}_p)$. Its role is played by the operator of
fractional differentiation $D^{\alpha}$ (the Vladimirov
pseudo-differential operator) which is defined as
\begin{equation}\label{a1}
D^{\alpha}f=\int_{\mathbb{Q}_p}|\xi|_p^{\alpha}F[f](\xi)\chi_p(-\xi{x})d_p\xi,
\qquad  \alpha>0.
\end{equation}

It is easy to see  that $D^{\alpha}f$ is well defined for all
$f\in\mathcal{D}(\mathbb{Q}_p)$. The element $D^{\alpha}f$ need not
belong necessarily to $\mathcal{D}(\mathbb{Q}_p)$ (since the
function $|\xi|_p^{\alpha}$ is not locally constant) however
$D^{\alpha}f\in{L}_2(\mathbb{Q}_p)$ \cite{KO}.

Since $\mathcal{D}(\mathbb{Q}_p)$ is not invariant with respect to
$D^{\alpha}$ we cannot define $D^{\alpha}$ on the whole space
$\mathcal{D}'(\mathbb{Q}_p)$. For a distribution
$f\in\mathcal{D}'(\mathbb{Q}_p)$ the operator $D^{\alpha}$ is well
defined only if the right-hand side of (\ref{a1}) exists\footnote{To
overcome such an inconvenience, a $p$-adic analog of the Lizorkin
spaces can be used instead of $\mathcal{D}(\mathbb{Q}_p)$
\cite{AKS}, \cite{AKS1}}.

In what follows we will consider $D^{\alpha}$, $\alpha>0$, as an
unbounded operator in $L_2(\mathbb{Q}_p)$. In this case, the domain
of definition $\mathcal{D}(D^{\alpha})$ consists of those
$f\in{L_2(\mathbb{Q}_p)}$ for which
$|\xi|_p^{\alpha}F[f](\xi)\in{L_2(\mathbb{Q}_p)}$. Since
$D^{\alpha}$ is unitarily equivalent to the operator of
multiplication by $|\xi|_p^{\alpha}$, this operator is positive
self-adjoint in $L_2(\mathbb{Q}_p)$ and its spectrum consists of
eigenvalues $\lambda_m=p^{\alpha{m}}$ $(m\in\mathbb{Z})$ of infinite
multiplicity and their accumulation point $\lambda=0$.

It was recently shown \cite{KOZ} that the set of eigenfunctions of
$D^{\alpha}$
\begin{equation}\label{a3}
\psi_{Nj\epsilon}(x)=p^{-\frac{N}{2}}\chi(p^{N-1}jx)\Omega(|p^{N}{x}-\epsilon|_p),
\quad N\in\mathbb{Z}, \ \epsilon\in\mathbb{Q}_p/\mathbb{Z}_p, \
j=1,\ldots,p-1
\end{equation}
forms an orthonormal basis in $L_2(\mathbb{Q}_p)$ ($p$-adic wavelet
basis) such that
\begin{equation}\label{a33}
D^{\alpha}\psi_{Nj\epsilon}=p^{\alpha(1-N)}\psi_{Nj\epsilon}.
\end{equation}
Here the indexes $N, j, \epsilon$ serve as parameters of the basis.
In particular, elements $\epsilon\in\mathbb{Q}_p/\mathbb{Z}_p$ can
be described as $\epsilon=\sum_{i=1}^{m}\epsilon_ip^{-i}$ \
($m\in\mathbb{N}, \ \epsilon_i=0,\ldots,p-1$).

The $p$-adic wavelet basis (\ref{a3}) does not depend on the choice
of $\alpha$ and it provides a convenient framework for the
investigation of $D^{\alpha}$. In particular, analyzing the
expansion of any element $u\in\mathcal{D}(D^{\alpha})$ with respect
to (\ref{a3}), it is not hard to establish the uniformly convergence
of the corresponding series for $\alpha>1/2$. This fact and the
property of eigenfunctions $\psi_{Nj\epsilon}$ to be continuous on
$\mathbb{Q}_p$ imply the next statement.

\begin{ppp}[\cite{KT}]\label{paf}
The domain $\mathcal{D}(D^{\alpha})$ consists of functions
continuous on $\mathbb{Q}_p$ if and only if $\alpha>1/2$.
\end{ppp}

Let us consider an equation
\begin{equation}\label{e12}
(D^{\alpha}-\lambda{I})h=\delta_{x_k}, \qquad
\lambda\in{\mathbb{C}}, \quad x_k\in\mathbb{Q}_p, \quad \alpha>0,
\end{equation}
where $D^{\alpha}: {L_2(\mathbb{Q}_p)} \to
\mathcal{D}'(\mathbb{Q}_p)$ is understood in the distribution sense.

It follows from \cite[Lemma 3.7]{KO} that  Eq. (\ref{e12}) has no
solutions belonging to $L_2(\mathbb{Q}_p)$ for $\alpha\leq{1/2}$.

\begin{ttt}\label{pifpaf}
The following statements are valid:

1. If $\alpha>1/2$, then  Eq. (\ref{e12}) has a unique solution
$h=h_{k,\lambda}\in{L_2(\mathbb{Q}_p)}$ if and only if
$\lambda\not=p^{\alpha{m}}$, where $m$ runs
$\mathbb{Z}\cup\{-\infty\}$.

2. If $\alpha>1$ and $\lambda\not=p^{\alpha{m}}$
($\forall{m}\in\mathbb{Z}\cup\{-\infty\}$), then
$h_{k,\lambda}\in\mathcal{D}(D^{\alpha/2})$.
\end{ttt}

{\it Proof.} First of all we remark that any function
$u\in\mathcal{D}(D^{\alpha})$ can be expanded in an uniformly
convergent series with respect to the complex-conjugated $p$-adic
wavelet basis $\{\overline{\psi_{Nj\epsilon}}\}$. This means (since
$\{\overline{\psi_{Nj\epsilon}}\}$ are continuous functions on
$\mathbb{Q}_p$) that
$u(x_k)=\sum_{N=-\infty}^{\infty}\sum_{j=1}^{p-1}\sum_{\epsilon}(u,\overline{\psi_{Nj\epsilon}})\overline{\psi_{Nj\epsilon}}(x_k)
$ for $x=x_k$.

Obviously, $\overline{\psi_{Nj\epsilon}}(x_k)\not=0 \iff
|p^Nx_k-\epsilon|_p\leq{1}$. Here
$\epsilon\in\mathbb{Q}_p/\mathbb{Z}_p$ and  hence, $|\epsilon|_p>1$
for $\epsilon\ne 0$. It follows from the strong triangle inequality
and the condition $\epsilon\in\mathbb{Q}_p/\mathbb{Z}_p$ that
$|p^Nx_k-\epsilon|_p\leq{1} \iff \epsilon=\{p^Nx_k\}_p$. But then,
recalling (\ref{a3}), we obtain
\begin{equation}\label{at67}
\overline{\psi_{Nj\epsilon}}(x_k)= \left\{ \begin{array}{ll}
0, &  \epsilon\ne\{p^Nx_k\}_p  \\
p^{-N/2}\chi(-p^{N-1}jx_k), &  \epsilon=\{p^Nx_k\}_p
\end{array}
\right.
\end{equation}
Therefore,
\begin{eqnarray}\label{oh1}
<\delta_{x_k},
u>=u(x_k) &=& \sum_{N=-\infty}^{\infty}\sum_{j=1}^{p-1}p^{-N/2}\chi(-p^{N-1}jx_k)\big(u,\overline{\psi_{Nj\{p^Nx_k\}_p}}\big) \\
&=&
\sum_{N=-\infty}^{\infty}\sum_{j=1}^{p-1}p^{-N/2}\chi(-p^{N-1}jx_k)<\psi_{Nj\{p^Nx_k\}_p},u>.
\nonumber
\end{eqnarray}

Since $\mathcal{D}(\mathbb{Q}_p)\subset\mathcal{D}(D^{\alpha})$ the
equality (\ref{oh1}) yields that
\begin{equation}\label{ee13}
\delta_{x_k}=\sum_{N=-\infty}^{\infty}\sum_{j=1}^{p-1}p^{-N/2}\chi(-p^{N-1}jx_k)\psi_{Nj\{p^Nx_k\}_p},
\end{equation}
where the series converges in  $\mathcal{D}'(\mathbb{Q}_p)$.

Suppose that a function $h\in{L_2(\mathbb{Q}_p)}$ is represented as
a convergent series in $L_2(\mathbb{Q}_p)$:
$$
h(x)=\sum_{N=-\infty}^{\infty}\sum_{j=1}^{p-1}\sum_{\epsilon}c_{Nj\epsilon}\psi_{Nj\epsilon}(x).
$$
Applying the operator $D^{\alpha}-\lambda{I}$ termwise, we get a
series
\begin{equation}\label{ee14}
(D^{\alpha}-\lambda{I})h=\sum_{N=-\infty}^{\infty}\sum_{j=1}^{p-1}\sum_{\epsilon}c_{Nj\epsilon}\big(p^{\alpha(1-N)}-\lambda\big)\psi_{Nj\epsilon},
\end{equation}
converging in $\mathcal{D}'$ (since
$D^{\alpha}\mathcal{D}(\mathbb{Q}_p)\subset{L_2}(\mathbb{Q}_p)$).
The comparison of (\ref{ee13}) and (\ref{ee14}) gives
$$
c_{Nj\epsilon}=\left\{
\begin{array}{ll}
  0, & \epsilon\ne\{p^Nx_k\}_p \\
  p^{-N/2}\chi(-p^{N-1}jx_k)\big[p^{\alpha(1-N)}-\lambda\big]^{-1}, &
  \epsilon=\{p^Nx_k\}_p
\end{array}
\right.
$$
Thus
\begin{equation}\label{ee15}
h_{k,
\lambda}(x)=\sum_{N=-\infty}^{\infty}\sum_{j=1}^{p-1}p^{-N/2}\chi(-p^{N-1}jx_k)\big[p^{\alpha(1-N)}-\lambda\big]^{-1}\psi_{Nj\{p^Nx_k\}_p}(x)
\end{equation}
is a unique solution of (\ref{e12}).

Since the functions $\psi_{Nj\{p^Nx_k\}_p}(x)$ in (\ref{ee15}) are
elements of the orthonormal basis  (\ref{a3}) in
$L_2(\mathbb{Q}_p)$,  the function $h_{k, \lambda}(x)$ belongs to
${L_2(\mathbb{Q}_p)}$ \ if and only if
$$
(p-1)\sum_{N=-\infty}^{\infty}p^{-N}\big[p^{\alpha(1-N)}-\lambda\big]^{-2}<\infty.
$$
 This inequality holds $\iff \ \lambda\not=p^{\alpha{m}}$
($\forall{m}\in\mathbb{Z}\cup\{-\infty\}$). Assertion 1 is proved.

Let $\alpha>1$. Taking (\ref{a3}) and (\ref{ee15}) into account, it
is easy to see that $h_{k,\lambda}\in\mathcal{D}(D^{\alpha/2})$ if
and only if the following series converge in $L_2(\mathbb{Q}_p)$:
\begin{eqnarray*}
{}& \sum_{N=1}^{\infty}\sum_{j=1}^{p-1}p^{-N/2}\chi(-p^{N-1}jx_k)\big[p^{\alpha(1-N)}-\lambda\big]^{-1}p^{\frac{\alpha}{2}(1-N)}\psi_{pj\{p^Nx_k\}_p}+\\
&\sum_{N=-\infty}^{0}\sum_{j=1}^{p-1}p^{-N/2}\chi(-p^{N-1}jx_k)\big[p^{\alpha(1-N)}-\lambda\big]^{-1}p^{\frac{\alpha}{2}(1-N)}\psi_{pj\{p^Nx_k\}_p}
\end{eqnarray*}
(if the limit exists then it coincides with $D^{\alpha/2}h_k$). For
the general term of the first series we have
$$
\big|p^{-N/2}p^{\frac{\alpha}{2}(1-N)}\chi(-p^{N-1}jx_k)\big[p^{\alpha(1-N)}-\lambda\big]^{-1}\big|^2\le
Cp^{-N(\alpha+1)}, \quad N\geq{1}
$$
(since $\lambda\not=p^{\alpha{m}}$,
$\forall{m}\in\mathbb{Z}\cup\{-\infty\}$) that implies its
convergence in $L_2(\mathbb{Q}_p)$ for $\alpha>1/2$.

Similarly, the general term of the second series can be estimated
from above by $Cp^{(\alpha-1)N}$  ($N\leq{0}$), which implies its
convergence in $L_2(\mathbb{Q}_p)$ for $\alpha>1$. Theorem
\ref{pifpaf} is proved. \rule{2mm}{2mm}

Let us study the solutions $h_{k, \lambda}(x)$ of (\ref{e12}) in
more detail for $\alpha>1$. To do this we consider the family of
functions $M_{p^{\gamma}}(\lambda)$ \
$(\gamma\in\mathbb{Z}\cup\{-\infty\})$ represented by the series
\begin{equation}\label{e32}
M_{p^{\gamma}}(\lambda)=\frac{p-1}{p}\sum_{N=-\infty}^{-\gamma}\frac{p^N}{p^{\alpha
N}-\lambda}-\frac{p^{-\gamma}}{p^{\alpha(1-\gamma)}-\lambda},\quad\gamma\in\mathbb{Z},
\end{equation}
\begin{equation}\label{e31}
M_{p^{-\infty}}(\lambda):=M_0(\lambda)=\frac{p-1}{p}\sum_{N=-\infty}^{\infty}\frac{p^N}{p^{\alpha
N}-\lambda}.
\end{equation}

Obviously, $M_0(\lambda)$ is differentiable for
$\lambda\in\mathbb{C}\setminus\{p^{\alpha{N}} | \
\forall{N}\in\mathbb{Z}\cup\{-\infty\}\}$ and
$M_0'(\lambda)=\frac{p-1}{p}\sum_{N=-\infty}^{\infty}\frac{p^N}{(p^{\alpha
N}-\lambda)^2}$.

\begin{ppp}\label{ah-oh}
Let $\alpha>1$ and $\lambda\not=p^{\alpha{N}}$
($\forall{N}\in\mathbb{Z}\cup\{-\infty\}$). Then
$$
h_{k, \lambda}(x)=\left\{\begin{array}{l} M_0(\lambda) \quad
\mbox{if} \quad x=x_k \\
M_{p^{\gamma}}(\lambda) \quad \mbox{if} \quad |x-x_k|_p=p^\gamma
\end{array}\right., \quad \|h_{k, \lambda}\|^2=M_0'(\lambda).
$$
\end{ppp}

{\it Proof.} If $\alpha>1$ and $\lambda\not=p^{\alpha{N}}$
($\forall{N}\in\mathbb{Z}\cup\{-\infty\}$), then
$h_{k,\lambda}\in\mathcal{D}(D^{\alpha/2})$, where $\alpha/2>1/2$
and hence, the series (\ref{ee15}) point-wise converges to
$h_{k,\lambda}(x)$.

Employing (\ref{a3}) and (\ref{e31}), we immediately deduce from
(\ref{ee15}) that \linebreak[4] $h_{k,\lambda}(x_k)=M_0(\lambda)$,
$\|h_{k, \lambda}\|^2=M_0'(\lambda)$, and
\begin{equation}\label{xixi}
h_{k,\lambda}(x)=\sum_{N=-\infty}^{\infty}\sum_{j=1}^{p-1}\frac{p^{-N}\chi(p^{N-1}j(x-x_k))}{p^{\alpha(1-N)}-\lambda}\cdot\Omega(|p^Nx-\{p^Nx_k\}_p|_p)
\end{equation}
for $x\not=x_k$.

The expression (\ref{xixi}) can be simplified with the use of the
following arguments: $1.$ It follows from the strong triangle
inequality and the definitions of $\{\cdot\}_p$ and $\Omega(\cdot)$
that $\Omega(|p^Nx-\{p^Nx_k\}_p|_p)=\Omega(|p^Nx-p^Nx_k|_p)$ and
$$
\Omega(|p^Nx-\{p^Nx_k\}_p|_p)=0\Leftrightarrow|p^N(x-x_k)|_p>
1\Leftrightarrow |x-x_k|_p>p^N.
$$
If $x\not=x_k$, then $|x-x_k|=p^{\gamma}$ for some
$\gamma\in\mathbb{Z}$. Therefore, the terms of (\ref{xixi}) with
indexes $N<\gamma$ are equal to zero.

$2.$ Since
$|p^{N-1}j(x-x_k)|_p=|p^{N-1}|_p|j|_p|x-x_k|_p=p^{\gamma+1-N}$ the
fractional part $\{p^{N-1}j(x-x_k)\}_p$ is equal to zero for
$N\geq{\gamma+1}$. Hence, $\chi(p^{N-1}j(x-x_k))\equiv{1}$ when
$N\geq{\gamma+1}$.

$3.$ Denote for brevity $y=p^{N-1}(x-x_k)$ and consider the case
when $N=\gamma$. Then $|y|_p=p$ and hence $\{y\}_p=p^{-1}y_0$, where
$y_0\in\{1,\ldots, p-1\}$ is a first term in the canonical
presentation of $y$ (see (\ref{e2})). Since $p$ is a prime number,
it is easy to verify that the set of numbers $\{jy\}_p$ $(j=1\dots
p-1)$ coincides with the set $p^{-1}j$ $(j=1\dots p-1)$ by modulo
$p$. This means that
$$
\sum_{j=1}^{p-1}\chi(p^{\gamma-1}j(x-x_k))=\sum_{j=1}^{p-1}\chi(jy)=\sum_{j=1}^{p-1}\exp\left(j\frac{2\pi
i}{p}\right)=-1
$$
(the last equality holds because $\sum_{j=1}^{p}\exp{ji\omega}=0$
for $\omega=\frac{2\pi}{p}$).

Statements 1.-3. allow one to rewrite (\ref{xixi}) as follows
$$
h_{k,\lambda}(x)=(p-1)\sum_{N=\gamma+1}^{\infty}\frac{p^{-N}}{p^{\alpha(1-N)}-\lambda}-\frac{p^{-\gamma}}{p^{\alpha(1-\gamma)}-\lambda}=M_{p^{\gamma}}(\lambda).
$$
Proposition \ref{ah-oh} is proved. \rule{2mm}{2mm}

By Proposition \ref{ah-oh},  $h_{k, \lambda}(x)$ is a `radial'
function which takes exactly one value $M_{p^{\gamma}}(\lambda)$ for
all points $x$ of the sphere $S_\gamma(x_k)=\{x\in\mathbb{Q}_p \ | \
|x-x_k|_p={p^\gamma}\}$. Such a property of the solution $h_{k,
\lambda}(x)$ of Eq. (\ref{e12}) is related to the property of
$\delta$ to be homogeneous of degree $|x|_p^{-1}$ \cite{VVZ}.

In conclusion, we single out properties of the functions
$M_{p^\gamma}(\lambda)$ and $M_0(\lambda)$  which will be useful for
the spectral analysis in the next section.
\begin{lelele}\label{dur1}
Let $\alpha>1$ and let $M_{p^\gamma}(\lambda)$ and $M_0(\lambda)$ be
defined by (\ref{e32}) and (\ref{e31}). Then:

$1.$ The function $M_0(\lambda)$ is continuous and monotonically
increasing on each interval $(-\infty, 0)$, \ $(p^{\alpha
N},p^{\alpha (N+1)}) \ (\forall{N}\in\mathbb{Z})$. Furthermore,
$M_0(\lambda)$ maps $(-\infty, 0)$ onto $(0, \infty)$ and maps
$(p^{\alpha N},p^{\alpha (N+1)})$ onto $(-\infty, \infty)$.

$2.$ The function $M_{p^\gamma}(\lambda)$ is continuous and
monotonically increasing (decreasing) on $(-\infty, 0)$ (on
$(p^{\alpha(1-\gamma)}, \infty)$). Furthermore,
$M_{p^\gamma}(\lambda)$ maps $(-\infty, 0)$ onto $(0, \infty)$ and
maps $(p^{\alpha(1-\gamma)}, \infty)$ onto $(0, \infty)$.
\end{lelele}

The proof of Lemma \ref{dur1} is quite elementary and it is based on
a simple analysis of the series (\ref{e32}) and (\ref{e31}). In
particular, rewriting the definition of $M_{p^\gamma}(\lambda)$ as
\begin{eqnarray*}
  M_{p^{\gamma}}(\lambda) &=& \sum_{N=-\infty}^{-\gamma}\frac{p^N}{p^{\alpha
N}-\lambda}-\sum_{N=-\infty}^{-\gamma+1}\frac{p^{N-1}}{p^{\alpha
N}-\lambda} \\
   &=& \sum_{N=-\infty}^{-\gamma}\frac{p^N}{p^{\alpha
N}-\lambda}-\sum_{N=-\infty}^{-\gamma}\frac{p^{N}}{p^{\alpha
(N+1)}-\lambda}=\sum_{N=-\infty}^{-\gamma}\frac{p^N(p^{\alpha(N+1)}-p^{\alpha
   N})}{(p^{\alpha
N}-\lambda)(p^{\alpha (N+1)}-\lambda)}
\end{eqnarray*}
we easy establish the assertion 2.

\setcounter{equation}{0}
\section{$p$-Adic Schr\"{o}dinger-Type Operator with Point Interactions}
In this section, we are going to study finite rank point
perturbations of $D^{\alpha}$ determined by the expression
\begin{equation}\label{eee1}
D^{\alpha}+V_Y, \qquad V_Y=\sum_{i,j=1}^nb_{ij}<\delta_{x_j},
\cdot>\delta_{x_i}, \quad   b_{ij}\in\mathbb{C}, \quad
Y=\{x_1,\ldots, x_n\}.
\end{equation}

Since $\delta_{x_j}\not\in{L}_2(\mathbb{Q}_p)$ the expression
(\ref{eee1}) does not determine an operator in $L_2(\mathbb{Q}_p)$.
Moreover, in contrast to the standard theory of point interactions
\cite{AL}, the potential $V_Y$ is not defined on the domain of
definition $\mathcal{D}(D^{\alpha})$ of the unperturbed operator
$D^{\alpha}$ for $\alpha\leq{1}/{2}$ (Proposition \ref{paf}). For
this reason we will assume $\alpha>{1}/{2}$.

\subsection{Definition of operator realizations of $D^{\alpha}+V$ in $L_2(\mathbb{Q}_p)$.}
Let $\mathfrak{H}_2\subset\mathfrak{H}_1\subset{L_2}(\mathbb{Q}_p)
 \subset\mathfrak{H}_{-1}\subset\mathfrak{H}_{-2}$
 be the standard scale of Hilbert spaces ($A$-scale) associated
with the positive self-adjoint operator $A=D^{\alpha}$ in
$L_2(\mathbb{Q}_p)$. Here ${\mathfrak H}_s=\mathcal{D}(A^{s/2})$,
$s=1,2$ with the norm $\|u\|_s=\|(D^{\alpha}+I)^{s/2}u\|$ and
${\mathfrak H}_{-s}$ is the completion of ${L_2}(\mathbb{Q}_p)$ with
respect to the norm $\|u\|_{-s}$. In a natural way ${\mathfrak
H}_{s}$ and ${\mathfrak H}_{-s}$ are dual and the inner product in
${L_2}(\mathbb{Q}_p)$ is extended to a pairing $ <\phi,
u>=((D^{\alpha}+I)^{s/2}u, (D^{\alpha}+I)^{-s/2}\phi),$ \
$u\in{{\mathfrak H}_{s}}, \phi\in{\mathfrak H}_{-s}$ (see \cite{AL1}
for details).

By virtue of Proposition \ref{ah-oh}, the solutions $h_{k,\lambda}$
of (\ref{e12}) satisfy the relation
$\overline{h_{k,\lambda}}=h_{k,\overline{\lambda}}$. Taking this
into account and using (\ref{oh1}) and (\ref{ee15}) we get
\begin{equation}\label{at4}
<\delta_{x_k},u>=u(x_k)=((D^{\alpha}-\overline{\lambda}{I})u, h_{k,
\lambda})_{L_2(\mathbb{Q}_p)} \quad (u\in{\mathcal{D}}(D^{\alpha}),
\ x_k\in\mathbb{Q}_p)
\end{equation}
for any complex $\lambda\not=p^{\alpha{m}}$
($\forall{m}\in\mathbb{Z}\cup\{-\infty\}$). Hence,
$\delta_{x_k}\in{\mathfrak H}_{-2}$.

In order to give a meaning to (\ref{eee1}) as an operator acting in
$L_2(\mathbb{Q}_p)$, we consider the positive symmetric operator
$A_{\mathrm{sym}}$ defined by:
\begin{equation}\label{ee4}
A_{\mathrm{sym}}={D^{\alpha}}\upharpoonright_{\mathcal{D}}, \quad
\mathcal{D}=\{u\in\mathcal{D}(D^{\alpha}) \ | \
u(x_1)=\ldots=u(x_n)=0\}, \quad \alpha>1/2.
\end{equation}

It follows from (\ref{at4}) that $A_{\mathrm{sym}}$ is a closed
densely defined symmetric operator in $L_2(\mathbb{Q}_p)$ and the
linear span of $\{h_{k, \lambda}\}_{k=1}^n$ coincides with
$\ker(A_{\mathrm{sym}}^*-\lambda{I})$. It is convenient to present
the domain of the adjoint ${\mathcal D}(A_{\mathrm{sym}}^*)$ as
${\mathcal
D}(A_{\mathrm{sym}}^*)={\mathcal{D}}(D^{\alpha})\dot{+}\mathcal{H}$,
where $\mathcal{H}=\ker(A_{\mathrm{sym}}^*+I)$. Then
\begin{equation}\label{ee5}
A_{\mathrm{sym}}^*f=A_{\mathrm{sym}}^*(u+h)=D^{\alpha}u-h, \qquad
\forall{f}=u+h\in{\mathcal D}(A_{\mathrm{sym}}^*)
\end{equation}
($u\in\mathcal{D}(D^{\alpha}), \ h\in\mathcal{H}$).

In the additive singular perturbation theory, the algorithm of the
determination of operator realizations of $D^{\alpha}+V_Y$ is well
known \cite{AL1} and it is based on the construction of some
extension (regularization)
${A}_{\mathrm{reg}}:=D^\alpha+V_{Y\mathrm{reg}}$ of (\ref{eee1})
onto the domain ${\mathcal
D}(A_{\mathrm{sym}}^*)={\mathcal{D}}(D^{\alpha})\dot{+}\mathcal{H}$.

The  $L_2(\mathbb{Q}_p)$-part
\begin{equation}\label{les40}
\widetilde{A}={A}_{\mathrm{reg}}\upharpoonright_{\mathcal{D}(\widetilde{A})},
\ \ \ \
\mathcal{D}(\widetilde{A})=\{f\in\mathcal{D}(A_{\mathrm{sym}}^*) \ |
\ {A}_{\mathrm{reg}}f\in{L_2(\mathbb{Q}_p)}\}
\end{equation}
of the regularization ${A}_{\mathrm{reg}}$ is called the {\it
operator realization} of $D^\alpha+V_Y$ in $L_2(\mathbb{Q}_p)$.

Since the action of $D^{\alpha}$ on elements of $\mathcal{H}$ is
defined by (\ref{e12}) the regularization ${A}_{\mathrm{reg}}$
depends on the definition of $V_{Y\mathrm{reg}}$.

If $\alpha>1$, Theorem \ref{pifpaf} gives that
$\delta_{x_k}\in\mathfrak{H}_{-1}$. Hence, the singular potential
$V_Y=\sum_{i,j=1}^nb_{ij}<\delta_{x_j}, \cdot>\delta_{x_i}$ is form
bounded \cite{AL}. In this case, the set
$\mathcal{D}(A_{\mathrm{sym}}^*)\subset\mathfrak{H}_{1}$ consists of
continuous functions on $\mathbb{Q}_p$ (in view of Proposition
\ref{paf} and Theorem \ref{pifpaf}) and  $\delta_{x_k}$ are uniquely
determined on elements $f\in\mathcal{D}(A_{\mathrm{sym}}^*)$ by the
formula (cf. (\ref{at4}))
\begin{equation}\label{as4}
<\delta_{x_k},f>=((D^{\alpha}+I)^{1/2}f,
(D^{\alpha}+I)^{1/2}{h_{k,-1}})_{L_2(\mathbb{Q}_p)}=f(x_k).
\end{equation}

Thus the regularization ${A}_{Y\mathrm{reg}}$ is uniquely defined
for $\alpha>1$ and formula (\ref{les40}) provides a unique operator
realization of (\ref{eee1}) in $L_2(\mathbb{Q}_p)$ corresponding to
a fixed singular potential $V_Y$.

If $1/2<\alpha\leq{1}$, then the delta functions $\delta_{x_k}$ form
a $\mathfrak{H}_{-1}$-independent system (since the linear span of
$\{\delta_{x_k}\}_1^n$ does not intersect with $\mathfrak{H}_{-1}$)
and $V_{Y\mathrm{reg}}$ is not uniquely defined on
$\mathcal{D}(A_{\mathrm{sym}}^*)$ (see \cite{KT} for a detailed
discussion of this part).

\subsection{Description of operator realizations.}

Let $\eta$ be an invertible bounded self-adjoint operator in
$L_2(\mathbb{Q}_p)$.

An operator $A$ is called $\eta$-{\it self-adjoint} in
$L_2(\mathbb{Q}_p)$ if $A^{*}={\eta}A{\eta^{-1}}$, where $A^{*}$
denotes the adjoint of $A$ \cite{AZ}. Obviously, self-adjoint
operators are $\eta$-self-adjoint ones for $\eta=I$. In that case we
will use the simpler terminology `self-adjoint' instead of
`$I$-self-adjoint'.

Our goal is to describe  $\eta$-self-adjoint operator realizations
of $D^{\alpha}+V_Y$ in $L_2(\mathbb{Q}_p)$ for $\alpha>1$.

Since the solutions $h_k:=h_{k,-1}$ ($1\leq{k}\leq{n}$) of
(\ref{e12}) form a basis of $\mathcal{H}$ any function
$f\in{\mathcal
D}(A_{\mathrm{sym}}^*)={\mathcal{D}}(D^{\alpha})\dot{+}\mathcal{H}$
admits a decomposition $f=u+\sum_{k=1}^n{c}_k{h}_k$ \ \
($u\in{\mathcal{D}}(D^{\alpha}), \  c_k\in\mathbb{C})$. Using such a
presentation we define the linear mappings $\Gamma_i :{\mathcal
D}(A_{\mathrm{sym}}^*)\to{\mathbb C}^n$ ($i=0,1$):
\begin{equation}\label{k9}
\Gamma_0f=\left(\begin{array}{c}
  f(x_1) \\
 \vdots \\
  f(x_n)
\end{array}\right), \quad \Gamma_1f=-\left(\begin{array}{c}
 c_1 \\
 \vdots \\
c_n
\end{array}\right), \quad
\forall{f}=u+\sum_{k=1}^n{c}_k{h}_k\in{\mathcal
D}(A_{\mathrm{sym}}^*).
\end{equation}

In what follows we assume that
\begin{equation}\label{ne2}
D^{\alpha}\eta={\eta}D^{\alpha} \qquad  \mbox{and} \qquad
 \eta:\mathcal{H}\to{\mathcal H}.
 \end{equation}

By the second relation in (\ref{ne2}), the action of $\eta$ on
elements of ${\mathcal H}$ can be described by the matrix
$\mathcal{Y}=\|{y}_{ij}\|^n_{i,j=1}$ where entries ${y}_{ij}$ are
determined by the relations $\eta{h_j}=\sum_{i=1}^n{y}_{ij}h_i$
($1\leq{j}\leq{n}$). In general, the basis $\{{h}_k\}_{i=1}^{n}$ of
$\mathcal{H}$ is not orthogonal and the matrix $\mathcal{Y}$ is not
Hermitian ($\mathcal{Y}\not=\overline{\mathcal{Y}}^{\mathrm{t}}$).

\begin{ttt}[\cite{KT}]\label{ss1}
Let $\alpha>1$ and let  $\widetilde{A}$ be the operator realization
of $D^{\alpha}+V_Y$ defined by (\ref{les40}). Then $\widetilde{A}$
coincides with the operator
\begin{equation}\label{k4141}
{A}_{\mathcal{B}}=A_{\mathrm{sym}}^*\upharpoonright_{\mathcal{D}(A_{\mathcal{B}})},
\quad
\mathcal{D}(A_{\mathcal{B}})=\{f\in\mathcal{D}(A_{\mathrm{sym}}^*) \
| \ {\mathcal B}\Gamma_0f=\Gamma_1f\},
\end{equation}
where ${\mathcal{B}}=\|b_{ij}\|_{i,j=1}^n$ is the coefficient matrix
of the potential $V_Y$.

The operator ${A}_{\mathcal{B}}$ is self-adjoint if and only if the
matrix $\mathcal{B}$ is Hermitian.

If $\eta$ satisfies (\ref{ne2}), then ${A}_{\mathcal{B}}$ is
$\eta$-self-adjoint if and only if the matrix $\mathcal{YB}$ is
Hermitian.
\end{ttt}

{\bf Example 1.} {\it $\mathcal{P}$-self-adjoint realizations.}

Let $Y=\{x_1, x_2\}$, where $x_2=-x_1$ and let $\eta=\mathcal{P}$ be
the space parity operator ${\mathcal P}f(x)=f(-x)$ in
$L_2(\mathbb{Q}_p)$. It follows from Proposition \ref{ah-oh} that
${\mathcal P}h_1=h_2$ and ${\mathcal P}h_2=h_1$. Hence, the
corresponding matrix $\mathcal{Y}$ has the form
$\mathcal{Y}=\left(\begin{array}{cc} 0 & 1 \\
1 & 0 \end{array}\right)$ and $\mathcal{P}$ satisfies (\ref{ne2}).

By Theorem \ref{ss1} the formula (\ref{k4141}) determines
$\mathcal{P}$-self-adjoint realizations ${A}_{\mathcal{B}}$ of
$D^{\alpha}+V_Y$ if and only if the entries $b_{ij}$ of the matrix
${\mathcal{B}}=\|b_{ij}\|_{i,j=1}^2$ satisfy the relations $b_{12},
b_{21}\in\mathbb{R}$, \ $b_{11}=\overline{b}_{22}$.

Under such conditions imposed on $b_{ij}$ the corresponding singular
potential $V_Y$ is not symmetric in the standard sense (except the
case $b_{ij}\in\mathbb{R}, \  b_{11}=b_{22}, \ b_{12}=b_{21}$) but
satisfies the condition of $\mathcal{P}$-symmetry
$\mathcal{P}V^*_Y=V_Y\mathcal{P}$, where the adjoint $V^*_Y$ is
determined by the relation $<V_Yu,v>=<u,V^*_Yv>$
($u,v\in\mathcal{D}(D^{\alpha})$). Assuming formally that
$\mathcal{T}V_Y=V^*_Y\mathcal{T}$, where $\mathcal{T}$ is the
complex conjugation operator $\mathcal{T}f(x)=\overline{f(x)}$, we
can rewrite the condition of $\mathcal{P}$-symmetry as follows
$\mathcal{PT}V_Y=V_Y\mathcal{PT}$. This means that the expression
$D^{\alpha}+V_Y$  is $\mathcal{PT}$-symmetric (since
$\mathcal{PT}D^{\alpha}=D^{\alpha}\mathcal{PT}$). Thus the
$\mathcal{P}$-self-adjoint operators ${A}_{\mathcal{B}}$ described
above are operator realizations of the $\mathcal{PT}$-symmetric
expression $D^{\alpha}+V_Y$ in $L_2(\mathbb{Q}_p)$.

\subsection{Spectral properties.}
As a rule, spectral properties of finite rank perturbations are
described in terms of a Nevanlinna function (Krein-Langer
$Q$-function) appearing as a parameter in a Krein's type resolvent
formula relating the resolvents of perturbed and unperturbed
operators \cite{AL1}, \cite{DHS}, \cite{P2}. The choice of a
resolvent formula has to be motivated by simple links with the
parameters of the perturbations.

Denote by $\mathcal{L}$ and $\mathcal{L}_Y$ the closed subspaces of
$L_2(\mathbb{Q}_p)$ spanned by the $p$-adic wavelets
$\psi_{Nj\epsilon}(x)$ $(N\in\mathbb{Z}, j=1,\ldots,p-1)$ with
$\epsilon\not=\{p^Nx_i\}_p$ $(\forall{x_i}\in{Y})$ and
$\epsilon=\{p^Nx_i\}_p$  $( \exists{x_i}\in{Y}\}$, respectively.
Obviously, $\mathcal{L}\oplus{\mathcal{L}_Y}=L_2(\mathbb{Q}_p)$.
Relations (\ref{a33}), (\ref{at67}), and (\ref{ee4}) imply that the
subspaces $\mathcal{L}$ and $\mathcal{L}_Y$ reduce the operators
$D^{\alpha}$ and $A_{\mathrm{sym}}$. Furthermore
$A_{\mathrm{sym}}=D^{\alpha}\upharpoonright_{\mathcal{L}}\oplus{A_{\mathrm{sym}}}\upharpoonright_{\mathcal{L}_Y}$.

Let $A_{\mathcal{B}}$ be the operator realization of $D^{\alpha}+V$
defined by (\ref{k4141}). Then
$A_{\mathcal{B}}=D^{\alpha}\upharpoonright_{\mathcal{L}}\oplus{A_{\mathcal{B}}}\upharpoonright_{\mathcal{L}_Y}$.
Therefore, the spectrum of $A_{\mathcal{B}}$ consists of eigenvalues
$\lambda=p^{\alpha{N}}$ ($\forall{N}\in\mathbb{Z}$) of infinite
multiplicity and their accumulation point $\lambda=0$.

To describe eigenvalues of finite multiplicity we consider the
matrix
\begin{equation}\label{as45}
M(\lambda)=\Big\| M_{|x_i-x_j|_p}(\lambda)\Big\|_{i,j=1}^n, \quad
\forall\lambda\in\mathbb{C}\setminus\{p^{\alpha{N}} \ | \
\forall{N}\in\mathbb{Z}\cup\{-\infty\}\},
\end{equation}
where the functions $M_{|x_i-x_j|_p}(\lambda)$
($|x_i-x_j|_p=p^{\gamma(x_i,x_j)}$) are defined by (\ref{e32}) and
(\ref{e31}).

\begin{ttt}\label{t55}
Let the matrix ${\mathcal{B}}$ in (\ref{k4141}) be invertible. Then
a point $\lambda\in\mathbb{C}\setminus\{p^{\alpha{N}} \ | \
\forall{N}\in\mathbb{Z}\cup\{-\infty\}\}$ is an eigenvalue of finite
multiplicity of $A_{\mathcal{B}}$ if and only if
$\det[M(\lambda)+\mathcal{B}^{-1}]=0$. In this case, the (geometric)
multiplicity of $\lambda$ is $n-r$, where $r$ is the rank of
$M(\lambda)+\mathcal{B}^{-1}$.

If $\det[M(\lambda)+\mathcal{B}^{-1}]\not=0$, then
$\lambda\in\rho(A_{\mathcal{B}})$ and the corresponding Krein's
resolvent formula has the form
\begin{equation}\label{at2}
(A_{\mathcal{B}}-\lambda{I})^{-1}=(D^{\alpha}-\lambda{I})^{-1}-(h_{1,\lambda},\ldots,
h_{n,\lambda})[M(\lambda)+\mathcal{B}^{-1}]^{-1}\left(\begin{array}{c}
(\cdot, h_{1,\overline{\lambda}}) \\
\vdots \\
(\cdot, h_{n,\overline{\lambda}})
\end{array}\right).
\end{equation}
\end{ttt}

{\it Proof.} It is easy to see from (\ref{ee15}) that $ h_{k,
\lambda}=u+h_{k,-1}$, where $u\in\mathcal{D}(D^{\alpha})$. This
relation and (\ref{k9}) give
\begin{equation}\label{at6}
\Gamma_1h_{k,\lambda}=(0, \ldots \underbrace{-1}_{k \ th}, \ldots
0)^{\mathrm{t}}.
\end{equation}

On the other hand, in view of Proposition \ref{ah-oh} and
(\ref{k9}),
\begin{equation}\label{at5}
\Gamma_0h_{k,\lambda}=(M_{|x_k-x_1|_p}(\lambda), \ldots
\underbrace{M_{0}(\lambda)}_{k \ th}, \ldots
M_{|x_k-x_n|_p}(\lambda))^{\mathrm{t}}.
\end{equation}

It is clear that $\lambda$ is an eigenvalue of finite multiplicity
of $A_{\mathcal{B}}$ if and only if $\lambda\not=p^{\alpha{N}}$
($\forall{N}\in\mathbb{Z}\cup\{-\infty\}$) and there exists a
nontrivial element
$f_\lambda\in\ker(A_{\mathrm{sym}}^*-\lambda{I})\cap\mathcal{D}(A_{\mathcal{B}})$.
Representing $f_\lambda$  as $f_\lambda=\sum_{k=1}^nc_kh_{k,
\lambda}$, using (\ref{as45}), (\ref{at6}), (\ref{at5}), and keeping
in mind that
$\mathcal{D}(A_{\mathcal{B}})=\ker(\Gamma_0-{\mathcal{B}^{-1}\Gamma_1})$,
we rewrite the latter condition as follows:
$[M(\lambda)+\mathcal{B}^{-1}](c_1,\ldots,c_n)^{\mathrm{t}}=0$.
Therefore, $\lambda$ is an eigenvalue if and only if this matrix
equation has a non-trivial solution. Obviously, the (geometric)
multiplicity of $\lambda$ is $n-r$, where $r$ is the rank of
$M(\lambda)+\mathcal{B}^{-1}$.

The resolvent formula (\ref{at2}) can be established by a direct
verification with the help of (\ref{at4}), (\ref{at6}), and
(\ref{at5}). Theorem \ref{t55} is proved. \rule{2mm}{2mm}

{\bf Remark.} It is easy to see that the triple $(\mathbb{C}^n,
-\Gamma_1, \Gamma_0)$, where $\Gamma_i$ are defined by (\ref{k9}) is
a boundary value space (BVS) of $A_{\mathrm{sym}}$ and the matrix
$M(\lambda)$ is the corresponding Weyl-Titchmarsh function of
$A_{\mathrm{sym}}$  \cite{DM}. From this point of view, Theorem
\ref{t55} is a direct consequence of the general BVS theory.
However, we prefer not to employ the general constructions in the
cases where the required results can be established in a more direct
way.

\subsection{The case of $\eta$-self-adjoint operator realizations.}
 One of the principal motivations for the study of $\eta$-self-adjoint
 operators in framework of the quantum mechanics is the observation that some of them
 have real spectrum (like self-adjoint operators) and, therefore, they can be used as
 alternates to standard Hamiltonians to explain experimental data \cite{MO2}.

 Since an arbitrary $\eta$-self-adjoint operator ${A}$ is
 self-adjoint with respect to the indefinite metric
 $[f,g]:=(\eta{f}, g), ({f,g}\in{L_2(\mathbb{Q}_p)})$
 one can attempt to develop a consistent quantum theory for
 $\eta$-self-adjoint Hamiltonians with real spectrum. However, in this
 case, we encounter the difficulty of dealing with a Hilbert space
 $L_2(\mathbb{Q}_p)$ equipped by the indefinite metric $[\cdot,\cdot]$.
 One of the natural ways to overcome this problem
 consists in the construction of a certain previously unnoticed physical symmetry
 $\mathcal{C}$ for ${A}$ (see, e.g., \cite{BBJ}, \cite{BT}, \cite{MO}).

By analogy with \cite{BBJ}, we will say that an $\eta$-self-adjoint
operator ${A}$ acting in $L_2(\mathbb{Q}_p)$ possesses the property
of $\mathcal{C}${\it -symmetry} if there exists a bounded linear
operator $\mathcal{C}$ in $L_2(\mathbb{Q}_p)$ such that the
following conditions are satisfied:

 $(i) \quad  A{\mathcal{C}}={\mathcal{C}}A;$ \vspace{2mm}

$(ii) \quad  {\mathcal{C}}^2=I;$ \vspace{2mm}

$(iii)$ \quad the sesquilinear form  $(f,
g)_{\mathcal{C}}:=[\mathcal{C}f,g] \ \
(\forall{f,g}\in{L_2(\mathbb{Q}_p)})$ determines an inner product in
$L_2(\mathbb{Q}_p)$  that is equivalent to initial one.
 \vspace{2mm}

 The existence of a ${\mathcal{C}}$-symmetry
for an $\eta$-self-adjoint operator $A$ ensures unitarity of the
dynamics generated by $A$ in the norm
$\|\cdot\|^2_{\mathcal{C}}=(\cdot,\cdot)_{\mathcal{C}}$.

In ordinary quantum theory, it is crucial that any state vector can
be expressed as a linear combination of the eigenstates of the
Hamiltonian. For this reason, it is natural to assume that every
physically acceptable $\eta$-self-adjoint operator must admit an
unconditional basis composed of its eigenvectors, or at least, of
its root vectors (see \cite{TT} for a detailed discussion of this
point).

\begin{ttt}\label{t45}
Let ${A}_{\mathcal{B}}$ be the $\eta$-self-adjoint operator
realization of $D^{\alpha}+V$ defined by (\ref{k4141}). Then the
following statements are equivalent:

(i) ${A}_{\mathcal{B}}$ possesses the property of
$\mathcal{C}$-symmetry;

(ii) the spectrum $\sigma({A}_{\mathcal{B}})$ is real and there
exists a Riesz basis of $L_2(\mathbb{Q}_p)$ composed of the
eigenfunctions of ${A}_{\mathcal{B}}$.
\end{ttt}

{\it Proof.}   It is known that the property of
${\mathcal{C}}$-symmetry for $\eta$-self-adjoint operators is
equivalent to their similarity to self-adjoint ones (\cite{AK2},
\cite{MO}). Hence, if ${A}_{\mathcal{B}}$ possesses
$\mathcal{C}$-symmetry, then there exists an invertible bounded
operator $Z$ such that
\begin{equation}\label{as67}
{A}_{\mathcal{B}}=ZHZ^{-1},
\end{equation}
where $H$ is a self-adjoint operator in $L_2(\mathbb{Q}_p)$. So, the
spectrum of ${A}_{\mathcal{B}}$ lies on the real axis. Furthermore,
it follows from Theorem \ref{t55} that $\sigma({A}_{\mathcal{B}})$
has no more than a countable set of points of condensations.
Obviously, this property holds for the spectrum of the self-adjoint
operator $H$. Applying now Lemma 4.2.7 in \cite{AZ}, we immediately
derive the existence of an orthonormal basis of $L_2(\mathbb{Q}_p)$
composed of the eigenfunctions of $H$. To complete the proof of the
implication $(i)\Rightarrow(ii)$ it is sufficient to use
(\ref{as67}).

Let us verify that $(ii)\Rightarrow(i)$. Indeed, if
$\{f_i\}_1^\infty$ is a Riesz basis composed of the eigenfunctions
of ${A}_{\mathcal{B}}$ (i.e., ${A}_{\mathcal{B}}f_i=\lambda_if_i, \
\lambda_i\in\mathbb{R}$), then $f_i=Ze_i$, where $\{e_i\}_1^\infty$
is an orthonormal basis of $L_2(\mathbb{Q}_p)$ and $Z$ is an
invertible bounded operator. This means that (\ref{as67}) holds for
a self-adjoint operator $H$ defined by the relations
$He_i=\lambda_ie_i$. Theorem \ref{t45} is proved. \rule{2mm}{2mm}

The next statement is a direct consequence of Theorem \ref{t45}.
\begin{cor}\label{C23}
An arbitrary self-adjoint operator realization ${A}_{\mathcal{B}}$
of $D^{\alpha}+V$ possesses a complete set of eigenfunctions in
$L_2(\mathbb{Q}_p)$.
\end{cor}

In conclusion we note that the spectral properties of
$\eta$-self-adjoint operators can have rather unexpected features.
In particular, the standard one-dimensional Schr\"{o}dinger operator
with a certain kind $\mathcal{PT}$-symmetric zero-range potentials
gives examples of $\mathcal{P}$-self-adjoint operators in
$L_2(\mathbb{R})$ whose spectra coincide with $\mathbb{C}$
\cite{AKD}.

\subsection{The Friedrichs extension.}

Let $A_F$ be the Friedrichs extension of the symmetric operator
$A_{\mathrm{sym}}$ defined by (\ref{ee4}). The standard arguments of
the extension theory lead to the conclusion (see \cite{KT} for
details) that $A_F=D^{\alpha}$ when $1/2<\alpha\leq{1}$ and
$$
A_F=A_{\mathrm{sym}}^*\upharpoonright_{\mathcal{D}(A_F)}, \quad
\mathcal{D}(A_F)=\{f(x)\in\mathcal{D}(A_{\mathrm{sym}}^*) \ | \
f(x_1)=\ldots={f(x_n)=0}\}
$$
when $\alpha>1$. In the latter case, $\mathcal{D}(A_F)=\ker\Gamma_0$
and the operator $A_F$ can formally be described by (\ref{k4141})
with $\mathcal{B}=\infty$.

Obviously, the essential spectrum of $A_F$ consists of the
eigenvalues $\lambda=p^{\alpha{N}}$ $(N\in\mathbb{Z})$ of infinite
multiplicity, and their accumulation point $\lambda=0$.

Let $\alpha>1$. Repeating step by step the proof of Theorem
\ref{t55} and taking the relation $\mathcal{D}(A_F)=\ker\Gamma_0$
into account, we conclude that the discrete spectrum
$\sigma_{\mathrm{dis}}(A_F)$ coincides with the set of solutions
$\lambda$ of the equation $\det{M(\lambda)}=0.$

The obtained relation allows one to establish some connections
between $\sigma_{\mathrm{dis}}(A_F)$ and the geometrical
characteristics of the set $Y$. To illustrate this fact we consider
the two points case $Y=\{x_1, x_2\}$.

Indeed, $\lambda\in\sigma_{\mathrm{dis}}(A_F)\iff$
$$
0=\det \Big\|
M_{|x_i-x_j|_p}(\lambda)\Big\|_{i,j=1}^2=(M_0(\lambda)-M_{p^{\gamma}}(\lambda))(M_0(\lambda)+M_{p^{\gamma}}(\lambda)),
$$
where $p^{\gamma}=|x_1-x_2|_p$. Therefore, the discrete spectrum is
determined by the equations $M_0(\lambda)-M_{p^{\gamma}}(\lambda)=0$
and $M_0(\lambda)+M_{p^{\gamma}}(\lambda)=0$.

In view of (\ref{e32}) and (\ref{e31}),
\begin{equation}\label{tpt1}
M_0(\lambda)-M_{p^{\gamma}}(\lambda)=\frac{p-1}{p}\sum_{N=-\gamma+2}^{\infty}\frac{p^N}{p^{\alpha
N}-\lambda}+\frac{p^{1-\gamma}}{p^{\alpha(1-\gamma)}-\lambda}.
\end{equation}

A simple analysis of (\ref{tpt1}) shows that the function
$M_0(\lambda)-M_{p^{\gamma}}(\lambda)$ is monotonically increasing
on the intervals $(-\infty, p^{\alpha(1-\gamma)})$ and $(p^{\alpha
N}, p^{\alpha (N+1)})$, $\forall{N}\ge -\gamma+1$ and it maps
$(-\infty, p^{\alpha(1-\gamma)})$ onto $(0,\infty)$ and maps
$(p^{\alpha N}, p^{\alpha (N+1)})$ onto $(-\infty, \infty)$. This
means that the set of solutions of
$M_0(\lambda)-M_{p^{\gamma}}(\lambda)=0$ coincides with the infinite
series of numbers  $\lambda=\lambda_{N}^-$, \ $N\ge -\gamma+1$ \
each of which is situated in the interval $(p^{\alpha N}, p^{\alpha
(N+1)})$. We will call the series of numbers
$\{\lambda_{N}^-\}_{N=-\gamma+1}^{\infty}$ {\it the type-$1$ part}
of the discrete spectrum of $A_F$. So, the type-$1$ part
$\sigma_{\mathrm{dis}}^{-}$ of $\sigma_{\mathrm{dis}}(A_F)$ consists
of solutions of the equation
$M_0(\lambda)-M_{p^{\gamma}}(\lambda)=0$.

By virtue of (\ref{e32}) and (\ref{e31}),
$M_0(\lambda)+M_{p^{\gamma}}(\lambda)=$
$$
=2\frac{p-1}{p}\sum_{N=-\infty}^{-\gamma}\frac{p^N}{p^{\alpha
N}-\lambda}+\frac{p-2}{p}\frac{p^{1-\gamma}}{p^{\alpha(1-\gamma)}-\lambda}+\frac{p-1}{p}\sum_{N=-\gamma+2}^{\infty}\frac{p^N}{p^{\alpha
N}-\lambda}.
$$
Analyzing this relation, it is easy to see that there exists exactly
one solution $\lambda=\lambda_{N}^+$ of
$M_0(\lambda)+M_{p^{\gamma}}(\lambda)=0$ lying inside an interval
$(p^{\alpha N}, p^{\alpha (N+1)})$ \ $\forall{N}\in\mathbb{Z}$. We
will call the infinite series of numbers
$\{\lambda_{N}^+\}_{-\infty}^{\infty}$ {\it the type-$2$ part}
$\sigma_{\mathrm{dis}}^{+}$ of the discrete spectrum
$\sigma_{\mathrm{dis}}(A_F)$.

Obviously,
$\sigma_{\mathrm{dis}}^{-}\cup\sigma_{\mathrm{dis}}^{+}=\sigma_{\mathrm{dis}}(A_F)$.
Let $N\ge -\gamma+1$ and let
$\lambda_{N}^{\pm}\in\sigma_{\mathrm{dis}}^{\pm}$ be the
corresponding discrete spectrum points in $(p^{\alpha N}, p^{\alpha
(N+1)})$. It follows from Lemma \ref{dur1} that
$\lambda_{N}^{+}<\lambda_{N}^{-}$. Therefore,
$\sigma_{\mathrm{dis}}^{-}\cap\sigma_{\mathrm{dis}}^{+}=\emptyset$.

Thus the discrete spectrum $\sigma_{\mathrm{dis}}(A_F)$ consists of
infinite series of eigenvalues of multiplicity one, which are
disposed as follows: an interval $(p^{\alpha N}, p^{\alpha(N+1)})$
contains exactly one eigenvalue $\lambda_{N}^{+}$ if $N<-\gamma$
(type-$2$ only) and exactly two eigenvalues
$\lambda_{N}^{+}<\lambda_{N}^{-}$ if $N\ge -\gamma+1$ (type-1 and
type-$2$).

The obtained description shows that the type-$1$ part
$\sigma_{\mathrm{dis}}^{-}$ of $\sigma_{\mathrm{dis}}(A_F)$ uniquely
determines the distance $|x_1-x_2|_p$.

In the general case $Y=\{x_1,\ldots, x_n\}$ the discrete spectrum
$\sigma_{\mathrm{dis}}(A_F)$ also contains the type-$1$ part.
Indeed, denote by $p^{\gamma_\mathrm{min}}$ the minimal distance
between the points of $Y$. Without loss of generality we may assume
that $|x_1-x_2|_p = p^{\gamma_\mathrm{min}}$. Then, by the strong
triangle inequality,
$|x_j-x_1|_p=|x_j-x_2|_p=p^{\gamma_j}\geq{p^{\gamma_\mathrm{min}}}$
for any point $x_j\in{Y}$, ($j\not=1,2$). This means that the first
two rows (columns) of the matrix $M(\lambda)$ (see (\ref{as45}))
differ from each other by the first two terms only. Subtracting the
second row from the first one we get
\begin{eqnarray*}
  \det M(\lambda) &=&  (M_0(\lambda)-M_{p^{\gamma_{\mathrm{min}}}}(\lambda))\left|\begin{array}{ccccc}
 1 & -1 & 0 & \dots & 0 \\
 M_{p^{\gamma_{\mathrm{min}}}}(\lambda) & M_0(\lambda) & M_{p^{\gamma_3}}(\lambda) & \dots & M_{p^{\gamma_n}}(\lambda) \\
 M_{p^{\gamma_3}}(\lambda) & M_{p^{\gamma_3}}(\lambda) & \ddots & & \\
 \vdots & \vdots & & \ddots & \\
 M_{p^{\gamma_n}}(\lambda) & M_{p^{\gamma_n}}(\lambda) & & & \ddots\\
\end{array}\right|.
\end{eqnarray*}

Thus the type-$1$ part $\sigma_{\mathrm{dis}}^{-}$ of the discrete
spectrum always exists and it characterizes the minimal distance
$p^{\gamma_{\mathrm{min}}}$ between elements of $Y$.

\subsection{Two points interaction.}

{\bf 1.} {\it Invariance with respect to the change of points of
interaction.} Let $Y=\{x_1, x_2\}$ and let the symmetric potential
$V_Y=\sum_{i,j=1}^2b_{ij}<\delta_{x_j}, \cdot>\delta_{x_i}$ be
invariant under the change $x_1\leftrightarrow{x_2}$. This means
that $b_{ij}\in\mathbb{R}$ and $b_{11}=b_{22}, \ b_{12}=b_{21}$. In
this case, the inverse $\mathcal{B}^{-1}$ of the coefficient matrix
$\mathcal{B}$ has the form
$\mathcal{B}^{-1}=\left(\begin{array}{cc} a & b \\
b & a \end{array}\right)$, where $a=b_{11}/\Delta$,
$b=-b_{12}/\Delta$, and $\Delta=b_{11}^2-b_{12}^2\not=0$. (We omit
the case $b_{11}=b_{12}=b_{21}=b_{22}$.)

The operator $A_{\mathcal{B}}$ is self-adjoint in
$L_2(\mathbb{Q}_p)$ and (by Theorem \ref{t55})
$$
\lambda\in\sigma_{\mathrm{dis}}(A_{\mathcal{B}})\iff(M_0(\lambda)-M_{p^{\gamma}}(\lambda)+a-b)(M_0(\lambda)+M_{p^{\gamma}}(\lambda)+a+b)=0,
$$
where $p^{\gamma}=|x_1-x_2|_p$. Thus, the description of
$\sigma_{\mathrm{dis}}(A_{\mathcal{B}})$ is similar to the
description of $\sigma_{\mathrm{dis}}(A_{F})$ and we can define some
analogs of the type-$1$
$$\sigma^{-}_{\mathrm{dis}}(A_{\mathcal{B}}):=\{\lambda\in\mathbb{R}\setminus\sigma(D^{\alpha})
\ | \ M_0(\lambda)-M_{p^{\gamma}}(\lambda)+a-b=0\}$$ and the
type-$2$ \
$\sigma^{+}_{\mathrm{dis}}(A_{\mathcal{B}}):=\{\lambda\in\mathbb{R}\setminus\sigma(D^{\alpha})
\ | \ M_0(\lambda)+M_{p^{\gamma}}(\lambda)+a+b=0\}$ parts of the
discrete spectrum $\sigma_{\mathrm{dis}}(A_{\mathcal{B}})$.

By analogy with the Friedrichs extension case (see (\ref{tpt1})),
$\sigma^{-}_{\mathrm{dis}}(A_{\mathcal{B}})$ contains an infinite
series of eigenvalues $\lambda_{N}^-$ lying in the intervals
$(p^{\alpha N}, p^{\alpha (N+1)})$ \ $\forall{N}\ge -\gamma+1$.
However, in contrast to the Friedrichs case, the interval $(-\infty,
p^{\alpha(-\gamma+1)})$ contains an additional (unique) point
$\lambda^-\in\sigma^{-}_{\mathrm{dis}}(A_{\mathcal{B}})$ if and only
if
$$
0<b-a \quad \mbox{and} \quad
b-a\not=[M_0(\lambda)-M_{p^{\gamma}}(\lambda)]|_{\lambda=p^{\alpha{m}}},
\quad -\infty\leq{m}\leq{-\gamma},
$$
where the difference
$[M_0(\lambda)-M_{p^{\gamma}}(\lambda)]|_{\lambda=p^{\alpha{m}}}$ is
determined by (\ref{tpt1}). In particular, $\lambda^-<0 \iff
0<b-a<M_0(0)-M_{p^{\gamma}}(0)=p^{(1-\alpha)(-\gamma+1)}.$

The type-$2$ part $\sigma^{+}_{\mathrm{dis}}(A_{\mathcal{B}})$
contains an infinite series of eigenvalues $\lambda_{N}^+$ lying in
the intervals $(p^{\alpha N}, p^{\alpha (N+1)})$ \
$\forall{N}\in\mathbb{Z}$ covering positive semi-axis. An additional
(unique) negative point
$\lambda^+\in\sigma^{+}_{\mathrm{dis}}(A_{\mathcal{B}})$ arises
$\iff$ $b+a<0$.

Obviously
$\sigma^{-}_{\mathrm{dis}}(A_{\mathcal{B}})\cup\sigma^{+}_{\mathrm{dis}}(A_{\mathcal{B}})=\sigma_{\mathrm{dis}}(A_{\mathcal{B}})$
but $\sigma^{-}_{\mathrm{dis}}(A_{\mathcal{B}})$ and
$\sigma^{+}_{\mathrm{dis}}(A_{\mathcal{B}})$ need not be disjoint.

{\bf 2.} {\it Examples of $\mathcal{P}$-self-adjoint realizations.}

Let $Y=\{x_1, x_2\}$, where $x_2=-x_1$ and let $A_{\mathcal{B}}$ be
$\mathcal{P}$-self-adjoint realizations of $D^{\alpha}+V_Y$
described in Example 1. We restrict ourselves to the case where the
inverse $\mathcal{B}^{-1}$ of the coefficient matrix $\mathcal{B}$
has the form
$\mathcal{B}^{-1}=\left(\begin{array}{cc} -ia & b \\
-b & ia \end{array}\right)$ ($a, b\in{\mathbb{R}}$).

The operator $A_{\mathcal{B}}$ is $\mathcal{P}$-self-adjoint in
$L_2(\mathbb{Q}_p)$ and $\lambda$ is an eigenvalue of finite
multiplicity of $A_{\mathcal{B}}$ if and only if
$$
(M_0(\lambda)-M_{p^{\gamma}}(\lambda))(M_0(\lambda)+M_{p^{\gamma}}(\lambda))+a^2+b^2=0
\qquad (p^{\gamma}=|2x_1|_p).
$$

Using properties of $M_0(\lambda)-M_{p^{\gamma}}(\lambda)$ and
$M_0(\lambda)+M_{p^{\gamma}}(\lambda)$ presented in Subsection 3.5,
it is easy to describe real eigenvalues of $A_{\mathcal{B}}$.
Precisely: $(i)$ The negative semiaxis $\mathbb{R}_-=(-\infty,0)$
belongs to $\rho(A_{\mathcal{B}})$. $(ii)$ If $N<-\gamma$, then the
interval $(p^{\alpha{N}}, p^{\alpha(N+1)})$ contains an eigenvalue
$\lambda_N$ of $A_{\mathcal{B}}$ such that
$p^{\alpha{N}}<\lambda_N<\lambda_N^+$, where $\lambda_N^+$ is the
corresponding type-$2$ discrete spectrum point of the Friedrichs
extension $A_F$. $(iii)$ If $N\ge -\gamma+1$, then eigenvalues of
$A_{\mathcal{B}}$ may appear only in the subinterval
$(\lambda_N^+,\lambda_N^-)\subset(p^{\alpha{N}},
p^{\alpha{(N+1)}})$, where $\lambda_N^-$ is the type-$1$ point of
$\sigma_{\mathrm{dis}}(A_F)$. Decreasing the parameters $a$ and $b$
we can guarantee the existence of such a type eigenvalues for a
fixed interval $(p^{\alpha{N}}, p^{\alpha{(N+1)}})$.

\subsection{One point interaction.}
Without loss of generality we will assume $x_1=0$. Then the general
expression (\ref{eee1}) takes the form
$D^{\alpha}+b<\delta_{0},\cdot>\delta_{0} \quad
(b\in\mathbb{R}\cup{\infty})$ and the corresponding self-adjoint
operator realizations $A_b$ in $L_2(\mathbb{Q}_p)$ are defined by
the formula
\begin{equation}\label{dur2}
A_bf=A_b(u+{\beta}h_{1,-1})=D^{\alpha}u-{\beta}h_{1,-1},
\end{equation}
where the parameter $\beta=\beta(u, b)\in\mathbb{C}$ is uniquely
determined by the relation $bu(0)=-\beta[1+bM_0(-1)]$. The operators
$A_b$ are self-adjoint extensions of the symmetric operator
$A_{\mathrm{sym}}={D^{\alpha}}\upharpoonright_{\mathcal{D}}, \quad
\mathcal{D}=\{u\in\mathcal{D}(D^{\alpha}) \ | \ u(0)=0\}$.

In our case, the subspace $\mathcal{L}_Y$ is the closed linear span
of $\psi_{Nj0}(x)$ $(N\in\mathbb{Z}, j=1,\ldots,p-1)$ and
$A_b=D^{\alpha}\upharpoonright_{\mathcal{L}}\oplus{A_b}\upharpoonright_{\mathcal{L}_Y}$.
The operator ${A_b}\upharpoonright_{\mathcal{L}_Y}$ is a
self-adjoint extension of
$A_{\mathrm{sym}}\upharpoonright_{\mathcal{L}_Y}$ and the points
$p^{\alpha{(1-N)}}$ are eigenvalues of multiplicity $p-2$ of the
symmetric operator
$A_{\mathrm{sym}}\upharpoonright_{\mathcal{L}_Y}$. The orthonormal
basis $\{\widetilde{\psi}_{Nj0}(x)\}_{j=1}^{p-2}$ of the
corresponding subspace
$\ker(A_{\mathrm{sym}}\upharpoonright_{\mathcal{L}_Y}-p^{\alpha{(1-N)}}I)$
can be chosen as follows:
\begin{equation}\label{ar2}
\widetilde{\psi}_{Nj0}(x)=\left(\frac{j}{j+1}\right)^{1/2}\left[\psi_{N(j+1)0}(x)-\frac{1}{j}\sum_{i=1}^{j}\psi_{Ni0}(x)\right]
\end{equation}

The decomposition
$A_b=D^{\alpha}\upharpoonright_{\mathcal{L}}\oplus{A_b}\upharpoonright_{\mathcal{L}_Y}$,
Lemma \ref{dur1}, and Theorem \ref{t55} allow one to describe in
detail the spectral properties of $A_b$ $(b\not=0)$. Precisely:

$(i)$ The operator $A_b$ is positive $\iff$ $b>0$. Otherwise
$(b<0)$, the unique solution of the equation $M_0(\lambda)=-1/b$ on
the semi-axis $(-\infty, 0)$ gives a negative eigenvalue
$\lambda_b^-$ of multiplicity one. The corresponding normalized
eigenfunction has the form
\begin{equation}\label{re1}
\phi_{b}^{-}(x)=\frac{h_{1,\lambda_b^-}(x)}{\sqrt{M'_0(\lambda_b^-)}}=\frac{1}{\sqrt{M'_0(\lambda_b^-)}}
\sum_{m=-\infty}^{\infty}\sum_{j=1}^{p-1}\frac{p^{-m/2}}{p^{\alpha(1-m)}-\lambda_b^-}\psi_{mj0}(x).
\end{equation}

$(ii)$ The positive part of the discrete spectrum of $A_b$ consists
of an infinite series of points $\lambda_{Nb}$ of multiplicity one,
each of which is the unique solution of $M_0(\lambda)=-1/b$ in the
interval $(p^{\alpha N},p^{\alpha (N+1)}) \ (N\in\mathbb{Z})$. The
corresponding normalized eigenfunction is (cf. (\ref{re1}))
\begin{equation}\label{r1}
\phi_{Nb}(x)=\frac{1}{\sqrt{M'_0(\lambda_{Nb})}}
\sum_{m=-\infty}^{\infty}\sum_{j=1}^{p-1}\frac{p^{-m/2}}{p^{\alpha(1-m)}-\lambda_{Nb}}\psi_{mj0}(x).
\end{equation}

$(iii)$ The points $p^{\alpha{(1-N)}}$ are eigenvalues of infinite
multiplicity of $A_b$. The orthonormal basis of the corresponding
subspace $\ker(A_b-p^{\alpha{(1-N)}}I)$ can be chosen as follows:
$$
\psi_{Nj\epsilon}(x) \quad (1\leq{j}\leq{p-1}, \quad
\epsilon\not=0), \qquad \widetilde{\psi}_{Nj0}(x) \quad
(1\leq{j}\leq{p-2}),
$$
where $\psi_{Nj\epsilon}(x)$ and $\widetilde{\psi}_{Nj0}(x)$ are
defined by (\ref{a3}) and (\ref{ar2}), respectively.

$(iv)$ The coefficient $b$ of the singular perturbation
$b<\delta_{0},\cdot>\delta_{0}$ is uniquely recovered by any point
of the discrete spectrum and
$$
\sigma_{\mathrm{dis}}(A_{b_1})\cap\sigma_{\mathrm{dis}}(A_{b_2})=\emptyset
\quad (b_1\not={b_2}); \qquad
\bigcup_{b\in\mathbb{R}}\sigma_{\mathrm{dis}}(A_{b})=\mathbb{R}\setminus\sigma(D^{\alpha}).
$$

Combining properties $(i)-(iii)$ with Corollary \ref{C23} we
immediately establish the following statement.
\begin{ppp}\label{p34}
The set of eigenfunctions of $A_b$
\begin{equation}\label{at69}
\begin{array}{l}
\psi_{Nj\epsilon}(x) \quad (N\in\mathbb{Z}, \quad
1\leq{j}\leq{p-1}, \quad \epsilon\not=0), \\
\widetilde{\psi}_{Nj0}(x) \quad  (N\in\mathbb{Z}, \quad
1\leq{j}\leq{p-2}), \\
\phi_{Nb}(x) \quad (N\in\mathbb{Z}), \\
\phi_{b}^{-}(x) \quad (\mbox{for the case} \ b<0 \ \mbox{only})
\end{array}
\end{equation}
forms an orthonormal basis of $L_2(\mathbb{Q}_p)$.
\end{ppp}

The Krein spectral shift
$\xi_b(\lambda)=\frac{1}{\pi}\arg[1+bM_0(\lambda+i0)]$ is easily
calculated
$$
\xi_b(\lambda)=\left\{\begin{array}{l} 0 \quad
\mbox{if} \quad \lambda\in(-\infty, \lambda_-)\bigcup[\bigcup_{-\infty}^{\infty}(\lambda_{N,b}, p^{\alpha(N+1)})] \\
1 \quad \mbox{if} \quad \lambda\in(\lambda_-,
0)\bigcup[\bigcup_{-\infty}^{\infty}(p^{\alpha{N}}, \lambda_{N,b})]
\end{array}\right.
$$
(the interval $(\lambda_-,0)$ is omitted for $b>0$). Therefore
\cite{SI},  the difference of the spectral projectors
$P_\lambda(A_b)-P_\lambda(D^\alpha)$
($P_\lambda:=P_{(-\infty,\lambda)}$) is trace class and $
\mathrm{Tr}[P_\lambda(A_b)-P_\lambda(D^\alpha)]=0$ for all
$\lambda\in\ker\xi_b(\lambda)$.

Let us consider the transformation of dilation
$Uf(x)=p^{-1/2}f(px)$. Obviously, $U$ is an unitary operator in
$L_2(\mathbb{Q}_p)$ and the $p$-adic wavelet basis
$\{\psi_{Nj\epsilon}(x)\}$ is invariant with respect to the dilation
\begin{equation}\label{aa22}
U\psi_{Nj\epsilon}(x)=\psi_{(N+1)j\epsilon}(x).
\end{equation}
Furthermore, in view of (\ref{a33})
\begin{equation}\label{as14}
U^mD^\alpha=p^{\alpha{m}}D^{\alpha}U^{m}, \qquad m\in\mathbb{Z}.
\end{equation}
In this sense the operator $D^{\alpha}$ is $p^{\alpha{m}}$-{\it
homogeneous} with respect to the one parameter family
$\mathfrak{U}=\{U^m\}_{m\in\mathbb{Z}}$ of unitary operators
\cite{AL1}, \cite{HK}.
\begin{ppp}\label{t7}
Among self-adjoint operators $A_b$ described by (\ref{dur2}) there
are only two $p^{\alpha{m}}$-homogeneous operators with respect to
the family $\mathfrak{U}$. One of them $A_0=D^{\alpha}$ is the
Krein-von Neumann extension of $A_{\mathrm{sym}}$, another one
coincides with the Friedrichs extension $A_{\infty}=A_F$.

An orthonormal basis of $L_2(\mathbb{Q}_p)$ composed of the
eigenfunctions of $A_b$ and invariant with respect to the dilation
$U$ exists if and only if $b=0$ or $b=\infty$.
\end{ppp}

{\it Proof.} The first part of the Proposition is a direct
consequence of \cite[subsection 4.4]{HK}.

The $p$-adic wavelet basis $\{\psi_{Nj\epsilon}(x)\}$ is an example
of an orthonormal basis composed of the eigenfunctions of $A_0$ and
invariant with respect to $U$.

Let us show that the orthonormal basis of eigenfunctions of
$A_\infty$ defined by (\ref{at69}) also is invariant with respect to
$U$. Indeed, relations (\ref{ar2}) and (\ref{aa22}) yield
$U\widetilde{\psi}_{Nj0}=\widetilde{\psi}_{(N+1)j0}$.

It follows from (\ref{e31}) that
\begin{equation}\label{asa1}
p^{\alpha-1}M_0(p^\alpha\lambda)=M_0(\lambda).
\end{equation}
Using (\ref{asa1}) and recalling that $\lambda_{N\infty}$ is the
solution of $M_0(\lambda)=0$ in the interval $(p^{\alpha
N},p^{\alpha (N+1)})$, we derive the recurrent relation
$\lambda_{(N+1)\infty}=p^\alpha\lambda_{N\infty}$. The obtained
relation and (\ref{r1}), (\ref{aa22}) imply
$U\phi_{N\infty}(x)=\phi_{(N-1)\infty}(x)$. Hence, the basis
(\ref{at69}) is invariant with respect to $U$ for $b=\infty$.

Let $\mathcal{M}$ be an arbitrary orthonormal basis composed of the
eigenfunctions of $A_b$ ($b\in\mathbb{R}\setminus\{0\}$). Since
$\lambda_{Nb}\in(p^{\alpha{N}}, p^{\alpha(N+1)})$ is an eigenvalue
of $A_b$ of multiplicity one the corresponding eigenfunction
$\phi_{Nb}(x)$ belongs to $\mathcal{M}$. Assuming that $\mathcal{M}$
is invariant with respect to $U$  we get
$A_bU\phi_{Nb}={\mu}U\phi_{Nb}$, where $\mu\in\sigma(A_b)$. To find
${\mu}$ we note that the $p^{\alpha{m}}$-homogeneity of $A_0$ and
$A_\infty$ with respect to $\mathfrak{U}$ implies that
$A_{\mathrm{sym}}$ and $A_{\mathrm{sym}}^*$ also are
$p^{\alpha{m}}$-homogeneous with respect to $U$. Therefore,
$$
\lambda_{Nb}U\phi_{Nb}=UA_b\phi_{Nb}=UA_{\mathrm{sym}}^*\phi_{Nb}=p^{\alpha}A_{\mathrm{sym}}^*U\phi_{Nb}=p^{\alpha}A_bU\phi_{Nb}=p^{\alpha}{\mu}U\phi_{Nb}.
$$
Thus $\mu=p^{-\alpha}\lambda_{Nb}$. Obviously
$\mu\in(p^{\alpha(N-1)}, p^{\alpha{N}})$ and $\mu$ is the solution
of $M_0(\lambda)=-1/b$ (since $\mu$ is an eigenvalue of $A_b$).
Employing (\ref{asa1}) for $\lambda=\mu$, we arrive at the following
contradiction $
-{1}/{b}=M_0(\mu)=p^{\alpha-1}M_0(p^\alpha\mu)=p^{\alpha-1}M_0(\lambda_{Nb})=-{p^{\alpha-1}}/{b}$
that completes the proof of Proposition \ref{t7}. \rule{2mm}{2mm}

\section{Acknowledgments} The
second (S.K.) and third (S.T.) authors thank DFG (project 436 UKR
113/88/0-1) and DFFD (project 14.01/003) for the financial support,
and the Institute f\"{u}r Angewandte Mathematik der Universit\"{a}t
Bonn for the warm hospitality.
 
 \end{document}